\documentclass[
reprint,
showkeys,
amsmath,amssymb,
aps,
PRR,
]{revtex4-2}

\usepackage{graphicx}
\usepackage{dcolumn}
\usepackage{bm}
\usepackage{hyperref}
\usepackage{pdfpages}
\usepackage{etoolbox} 
\makeatletter
\patchcmd{\@outputpage@head}{\@ifx{\LS@rot\@undefined}{}{\LS@rot}}{}{}{}
\makeatother

\begin{document}


\title{Suppression of chaos in a partially driven recurrent neural network}

\author{Shotaro Takasu}
 \email{shotaro.takasu.63x@st.kyoto-u.ac.jp}
\author{Toshio Aoyagi}%
\affiliation{%
  Graduate School of Informatics, Kyoto University, Yoshida-Honmachi, Sakyo-ku, Kyoto 606-8501, Japan
}%

\date{\today}

\begin{abstract}
The dynamics of recurrent neural networks (RNNs), and particularly their response to inputs, play a critical role in information processing. In many applications of RNNs, only a specific subset of the neurons generally receive inputs. However, it remains to be theoretically clarified how the restriction of the input to a specific subset of neurons affects the network dynamics. Considering RNNs with such restricted input, we investigate how the proportion, $p$, of the neurons receiving inputs (the "inputs neurons") and the strength of the input signals affect the dynamics by analytically deriving the conditional maximum Lyapunov exponent. Our results show that for sufficiently large $p$, the maximum Lyapunov exponent decreases monotonically as a function of the input strength, indicating the suppression of chaos, but if $p$ is smaller than a critical threshold, $p_c$, even significantly amplified inputs cannot suppress spontaneous chaotic dynamics. Furthermore, although the value of $p_c$ is seemingly dependent on several model parameters, such as the sparseness and strength of recurrent connections, it is proved to be intrinsically determined solely by the strength of chaos in spontaneous activity of the RNN. This is to say, despite changes in these model parameters, it is possible to represent the value of $p_c$ as a common invariant function by appropriately scaling these parameters to yield the same strength of spontaneous chaos. Our study suggests that if $p$ is above $p_c$, we can bring the neural network to the edge of chaos, thereby maximizing its information processing capacity, by amplifying inputs.

\end{abstract}

\maketitle
\section{INTRODUCTION}
Large-scale recurrent neural networks (RNNs) exhibit various dynamical patterns, including limit cycles and chaos. They have been used to model brain functions such as working memory \cite{Rajan2016}, motor control \cite{Laje2013}, and context-based learning \cite{Enel2016}. The rich dynamics exhibited by an RNN can also be used for information processing. Reservoir computing (RC) \cite{Jaeger2001a,Maass2002} is a machine learning framework that utilizes large RNNs, called “reservoirs”, to reproduce the time series data of interest. RC is not restricted to RNNs, and indeed a wide range of dynamical systems can serve as the reservoir under appropriate conditions. Physical reservoir computing, in which a real physical system is used as the reservoir, has been an area of active research in recent years \cite{Nakajima}. 

In general, RNNs must possess rich dynamics in order to assimilate a diverse set of signals to be learned. For this reason, it is advantageous for RNNs to exhibit chaotic spontaneous activity. On the other hand, for an RNN to successfully reproduce a target time series, it must converge to the same state each time it receives a particular set of input signals, regardless of its initial internal state.  This property is known as the "echo state property" \cite{Jaeger2001a} in the context of RC. Hence, it is hypothesized that an RNN that displays varied spontaneous activity while maintaining consistency in response to inputs will exhibit superior computational performance. Such an RNN is commonly referred to as being at the "edge of chaos", and it is known empirically that reservoirs in this regime have the highest computational capacity \cite{Bertschinger2004}. In fact, there is experimental evidence suggesting that mammalian neuronal networks operate in this critical regime \cite{Morales2023, Dahmen2019}.

Lyapunov spectrum analysis \cite{Pikovsky2016} allows us to study the dynamics of RNNs in a quantitative manner. The maximum Lyapunov exponent (MLE) characterizes the exponential rate of separation of infinitesimally close trajectories. In the case that a dynamical system is driven by given input signals, the MLE is called the maximum conditional Lyapunov exponent (MCLE). Two identical RNNs with slightly different initial states will converge to the same state under the same inputs if and only if the MCLE is negative. Therefore, a negative MCLE is necessary for an RNN to exhibit consistency with respect to inputs.

It is known that the MCLE of a random RNN decreases with the strength of the input signal and eventually becomes negative \cite{Molgedey1992,Massar2013,Haruna2019,Schuecker2018, Rajan2010}. In other words, sufficiently amplified driving input signals can suppress chaotic dynamics. These findings suggest that it is possible to shift the state of an RNN exhibiting chaotic spontaneous dynamics toward the edge of chaos by appropriately amplifying the input, and then use its shifted state as an efficient reservoir \cite{Sussilo2009}. There is also experimental evidence indicating similar suppression of chaos in the brain \cite{Churchland2010}.

Previous studies of the Lyapunov exponents of large random RNNs assume a model in which every unit in the RNN connects to the input layer and receives driving signals. Hereafter, we refer to RNNs of this type as "full-input RNNs". However, such a model is biologically implausible because biological synapses are sparse \cite{Wildenberg2021}. Additionally, in physical reservoir computing, it is often unfeasible to connect the input to all reservoir units. Therefore, it is important to investigate the dynamics of RNNs in the case that only a subset of the neurons receive input signals. We refer to RNNs of this type as "partial-input RNNs.” It remains to be elucidated whether sufficiently amplified inputs always suppress the chaotic activity in a partial-input RNN, as in the case of a full-input RNN. In this work, we address this question by analytically calculating the MCLE of a partial-input random RNN.

\section{Network Model}

\begin{figure}
    \centering
    \includegraphics[width=0.9\linewidth]{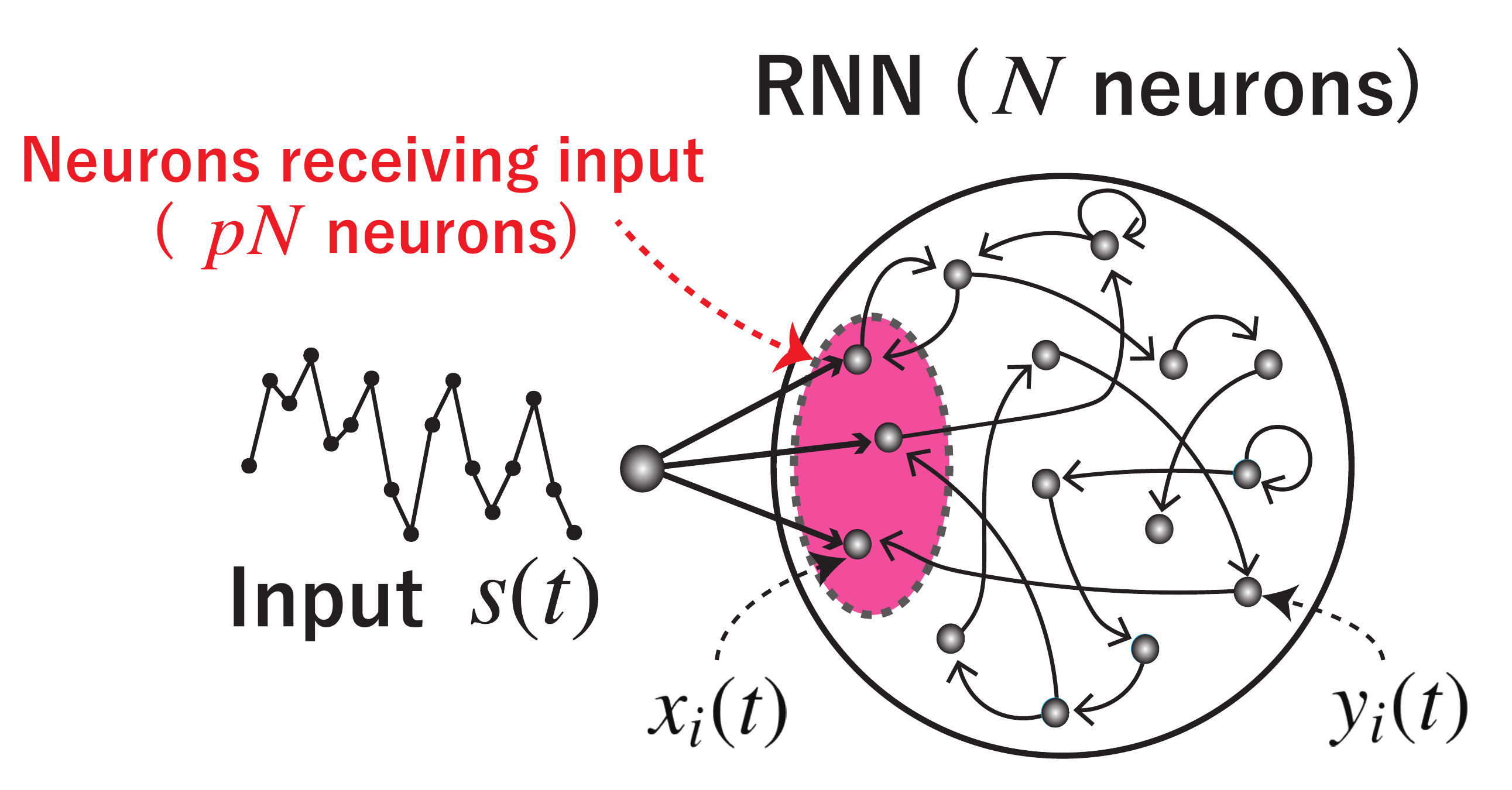}
    \caption{
        A schematic depiction of the partial-input RNN studied in this work. The shaded region represents the neurons that receive input signals through input connectivity.
    }
    \label{reservoir_schematic}
\end{figure}

We investigate the discrete-time dynamics of a random sparse RNN with $N \gg 1$ neurons of which only $pN$ receive inputs (Fig.\ref{reservoir_schematic}). We define the parameter $p\in [0,1]$, called the "input partiality," which determines the fraction of neurons coupling to the input unit. We consider the case in which the input signal $s(t)$ at time $t$ is a scalar for simplicity, although our theory can be straightforwardly extended to multi-dimensional inputs. The state of a neuron that receives inputs is represented by a dynamical variable $x_i(t)\in \mathbb{R}$ ($i=1,\cdots,pN$) whose evolution obeys the equation
\begin{eqnarray}
\label{x_dynamics}
x_{i}(t+1) = \sum_{j=1}^{pN}  J_{ij}\phi(x_{j}(t)) + \sum_{k=pN+1}^{(1-p)N}  J_{ik}\phi(y_{k}(t)) + u_i s(t),\nonumber \\
\end{eqnarray}
while the state of a neuron that does not receive inputs is represented by a dynamical variable $y_i(t)\in \mathbb{R}$ ($i=pN+1,\cdots,N$) whose evolution obeys the equation
\begin{eqnarray}
\label{y_dynamics}
y_{i}(t+1) &&= \sum_{j=1}^{pN}  J_{ij}\phi(x_j(t)) + \sum_{k=pN+1}^{(1-p)N}  J_{ik}\phi(y_k(t)). 
\end{eqnarray}
Here, $u_i$ ($i=1,\cdots,pN$) is the coupling weight connecting the input signal to the $i$th neuron, $J_{ij}$ is a recurrent weight matrix determining the coupling from the $j$th to the $i$th neuron, and $\phi$ is the activation function. The value of $u_i$ is drawn randomly from a Gaussian distribution with zero mean and unit variance. We define $J_{ij}$ to be a random sparse matrix whose elements are non-zero with probability $\alpha \in (0,1]$. The non-zero element is independently drawn from a Gaussian distribution with zero mean and variance $g^2/N$. The gain parameter $g$ represents the recurrent coupling strength. The activation function is chosen as $\phi(x) = \textrm{erf}(\frac{\sqrt{\pi}}{2}x)$ for analytic tractability. It has been found that in the absence of inputs, this RNN exhibits a transition from fixed-point dynamics to chaotic dynamics at $g=1$ in the limit of a large network \cite{Sompolinsky1988}. In previous studies, zero mean white Gaussian noise has typically been used for the input signal $s(t)$ \cite{Molgedey1992,Massar2013,Haruna2019,Schuecker2018}, but we do not limit the choice of $s(t)$ in this way, and instead regard it to be a time series that satisfies a certain condition to be shown in Sec.\ref{section lambda infty} . Overall, the model parameters to be varied are the input partiality, $p$, the recurrent connection sparsity, $\alpha$, and the recurrent coupling strength, $g$.

\section{Derivation of the maximum conditional Lyapunov exponent}

\subsection{Mean-field theory}

We can obtain the statistical properties of $x_i(t)$ and $y_i(t)$ using a mean-field approach \cite{Massar2013,Haruna2019} in the limit $N \rightarrow \infty$. As seen from Eqs.(\ref{x_dynamics}) and (\ref{y_dynamics}), $x_i(t+1)$ and $y_i(t+1)$ are sums of large numbers of identically distributed independent variables, $\{ J_{ij} \phi(x_j(t)) \}_{j=1}^{pN}$ and $\{ J_{ik} \phi(y_k(t)) \}_{k=pN+1}^{N}$. Thus, according to the central limit theorem, we can consider $x_i(t)$ and $y_i(t)$ to follow Gaussian distributions. It thus suffices to determine their averages, $\langle x_i(t) \rangle$ and $\langle y_i(t) \rangle$, and variances, $\langle x_i^2(t) \rangle$ and $\langle y_i^2(t) \rangle$, where $\langle \cdots \rangle$ denotes the average over realizations of the quenched weights $J_{ij}$ and $u_i$. 

Taking the average of the evolution equation of $x_i(t)$ and $x_i(t)^2$ over realizations of weights $J_{ij}$ and $u_i$, we obtain
\begin{eqnarray}
\label{x_mean}
\langle x_i(t+1) \rangle &&= 0, \nonumber \\
\label{x_variance}
\langle x_i(t+1)^2 \rangle  &&=  p\alpha g^2 \big \langle \phi(x_i(t))^2 \big \rangle \nonumber \\
        && \hspace{35pt} + (1-p)\alpha g^2 \big \langle \phi(y_k(t))^2 \big \rangle + s(t)^2. \nonumber
\end{eqnarray}
Similarly, $\langle y_i(t+1) \rangle$ and $\langle y_i(t+1)^2 \rangle$ are given by  
\begin{subequations}
\begin{eqnarray}
\label{y_mean}
\langle y_i (t+1) \rangle &&= 0,  \\
\label{y_variance}
\langle y_i(t+1)^2 \rangle  &&= p\alpha g^2 \big \langle \phi(x_i(t))^2 \big \rangle \nonumber \\
            && \hspace{50pt} + (1-p)\alpha g^2 \big \langle  \phi(y_k(t))^2 \big \rangle. 
\end{eqnarray}
\end{subequations}
Note that we have used the assumption that $\phi(x_j)$ and $\phi(y_k)$ is independent of its incoming weight $J_{ij}$ and $J_{ik}$. This assumption is justified in the limit $N \to \infty$ using the generating-function formalism \cite{Molgedey1992,Toyoizumi2011}.

Introducing the notation $K(t):=\langle y_i(t)^2 \rangle$, the Gaussian distributions exhibited by $x_i(t)$ and $y_i(t)$ can be expressed as follows:
\begin{subequations}
\label{xy_Gaussian}
\begin{eqnarray}
x_i(t+1) &&\overset{\rm{i.i.d}}{\sim} \mathcal{N}(0,K(t+1)+s(t)^2),\\
y_i(t+1) &&\overset{\rm{i.i.d}}{\sim} \mathcal{N}(0,K(t+1)).
\end{eqnarray}
\end{subequations}
From Eqs.(\ref{y_variance}) and (\ref{xy_Gaussian}), we can directly calculate $K(t+1)$, obtaining
\begin{widetext}
\begin{eqnarray}
\label{selfconsistentEq}
K(t+1) &&= p\alpha g^2\int Dx \phi^2 \left( \sqrt{K(t)+s(t-1)^2}x \right) + (1-p)\alpha g^2 \int Dy \phi^2 \left(\sqrt{K(t)}y \right) \nonumber \\
    &&= -\alpha g^2+\dfrac{4}{\pi} \alpha g^2 \left((1-p)\arctan{\sqrt{1+\pi K(t)}}+p \arctan \sqrt{1+\pi (K(t)+s(t-1)^2)} \right),
\end{eqnarray}
\end{widetext}

where $\int Dx := \int dx \dfrac{1}{\sqrt{2\pi}}e^{-\frac{x^2}{2}}$.

\subsection{Derivation of the MCLE}
The time series $\{ K(t) \}_t $ obtained above allows us to derive the MCLE, $\lambda$, of the RNN \cite{Molgedey1992}. The MCLE is defined as the asymptotic growth rate of the distance between two replicated RNNs, 
\begin{eqnarray}
\lambda &&:= \lim_{T \to \infty, \bm{\delta}(0)\to 0} \dfrac{1}{T} 
            \log 
            \dfrac{\| \bm{\delta}(T) \|}{\| \bm{\delta}(0) \|}, 
\label{MCLE_define}
\end{eqnarray}
where $\| \bm{\delta}(t) \|$ denotes the distance at time $t$ between two replicated RNNs receiving the same input signals. 

Linearizing the evolution equation of the RNN, we obtain the variational equation describing the evolution of infinitesimal perturbation $\bm{\delta}(t)$,
\begin{eqnarray}
\bm{\delta}(t+1) = \bm{J} \bm{\Phi}'(t) \bm{\delta}(t), \nonumber
\end{eqnarray}
where the matrix $\Phi'(t)$ is the diagonal matrix whose $i$th diagonal entry is $\phi'(x_i(t))$ for $1\leq i \leq pN $ or $\phi'(y_i(t))$ for $pN+1\leq i \leq N $. The typical growth rate, \\$\| \bm{\delta}(t+1) \| / \| \bm{\delta}(t) \|$, is determined by the spectral radius of the Jacobian $\bm{J}\Phi'(t)$. According to the random matrix theory \cite{Ahmadian2015}, the spectral radius $\rho(t)$ is given by
\begin{eqnarray}
\rho(t)^2 = p\alpha g^2 \langle \phi'(x(t))^2 \rangle + (1-p)\alpha g^2 \langle \phi'(y(t))^2 \rangle, \nonumber
\end{eqnarray}
in the limit of large network size. Applying the results of the mean-field theory (Eqs.(\ref{xy_Gaussian})), we can calculate $\lambda$, yielding 
\begin{widetext}
\begin{eqnarray}
\lambda &&= \lim_{T \to \infty, \bm{\delta}(0)\to 0} \dfrac{1}{2T} 
            \sum_{t=0}^{T-1} \log
            \dfrac{\| \bm{\delta}(t+1) \|^2}{\| \bm{\delta}(t) \|^2} 
         = \lim_{T \to \infty, \bm{\delta}(0)\to 0} \dfrac{1}{2T} 
            \sum_{t=0}^{T-1} \log \rho(t)^2
            \nonumber \\
        &&= \lim_{T \to \infty} \dfrac{1}{2T} 
            \sum_{t=0}^{T-1} \log
            \left(
            p\alpha g^2\int Dx \phi'^2 \left( \sqrt{K(t)+s(t-1)^2}x \right) + (1-p)\alpha g^2 \int Dy \phi'^2 \left(\sqrt{K(t)}y \right) 
            \right) \nonumber \\
        &&= \lim_{T \to \infty} \dfrac{1}{T}\sum_{t=0}^T \dfrac{1}{2} \log \alpha g^2 \left(\dfrac{p}{\sqrt{1+\pi(K(t)+s(t-1)^2)}} + \dfrac{1-p}{\sqrt{1+\pi K(t)}}
            \right).
\label{MCLE}
\end{eqnarray}
\end{widetext}

\begin{figure}[t]
    \centering
    \includegraphics[width=0.9\linewidth]{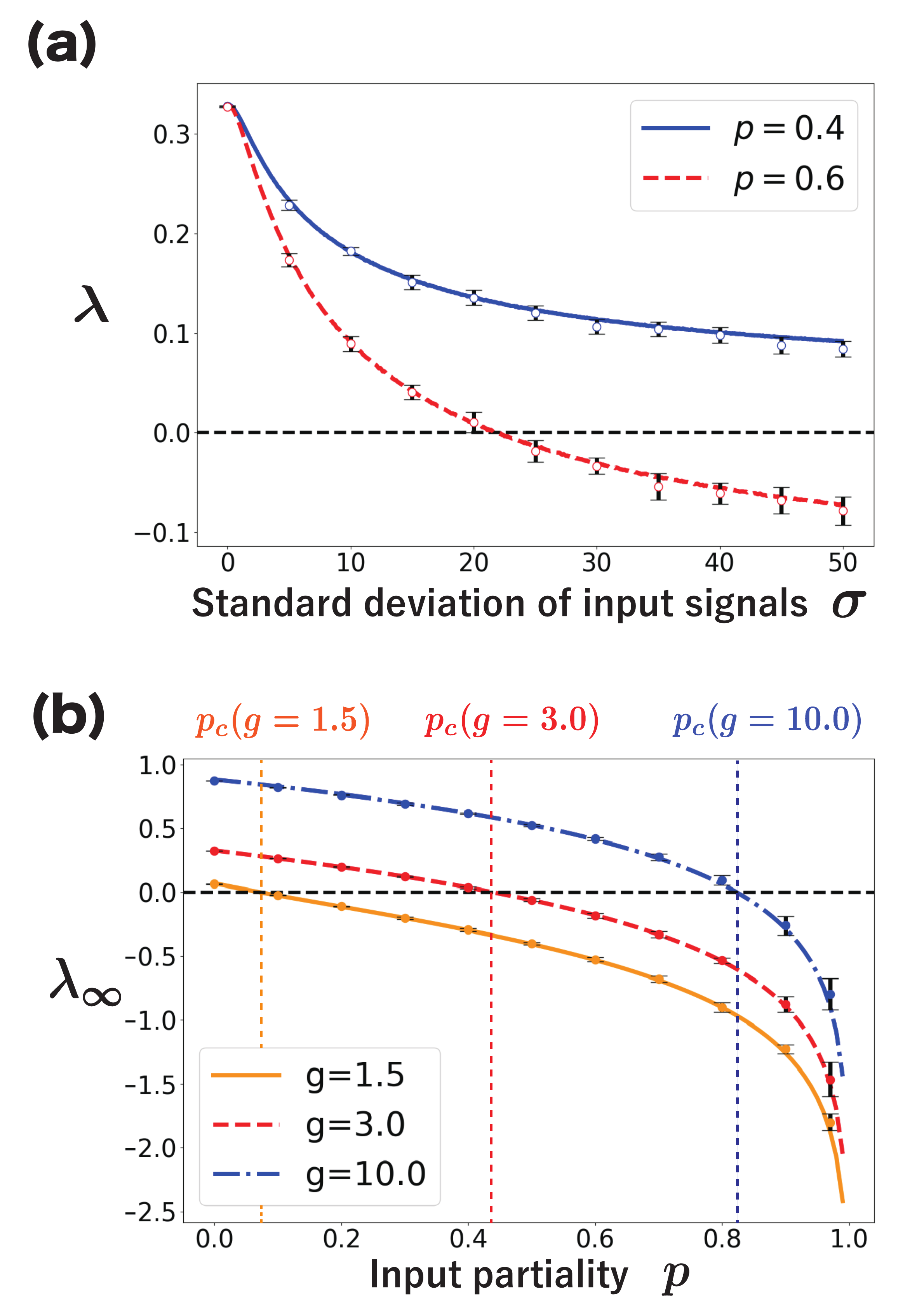}
    \caption{
    (a) The maximum conditional Lyapunov exponent (MCLE), $\lambda$, calculated for various values of the input partiality, $p$. The theoretical form given in Eq.(\ref{MCLE}) is plotted with a solid curve for $p=0.4$ and a dashed curve for $p=0.6$, where the time average in Eq.(\ref{MCLE}) is computed up to $T=10^5$. Error bars represent $\pm$std of direct numerical simulations based on the definition of $\lambda$ in Eq.(\ref{MCLE_define}). For each plot, the coupling strength is set to $g=3.0$, and the sparsity is set to $\alpha=1.0$. For the numerical simulations, the value of the MCLE was obtained by directly calculating Eq.(\ref{MCLE_define}) for a sufficiently long time, i.e., $T=10^4$. 
    (b) Analytic results (solid, dashed, dot-dashed curves) and numerical results (error bars indicating $\pm$std) for $\lambda_{\infty}$ with sparsity $\alpha=1.0$. In numerical simulations, the value of $\lambda_\infty$ was obtained by directly calculating Eq.(\ref{MCLE_define}) for a sufficiently large input magnitude ($\sigma=10^3$). In the cases of both (a) and (b), the input is assumed to be Gaussian white noise with zero mean and standard deviation $\sigma$. The values of the MCLE ($\lambda$ and $\lambda_\infty$) were calculated for 10 different network realizations with a network size of $N=1000$. 
    }
    \label{fig:lambda}
\end{figure}

Figure \ref{fig:lambda}(a) displays the MCLE, $\lambda$, as a function of the input intensity $\sigma$ defined below, for $p=0.4$ and $p=0.6$, with $g=3.0$ and $\alpha=1.0$. Here we assume the input to be Gaussian white noise with zero mean and standard deviation $\sigma$. The analytic results obtained from Eq.(\ref{MCLE}) are consistent with the results of the numerical simulations. For both values of $p$, $\lambda$ decreases monotonically as a function of $\sigma$. However, while the MCLE for $p=0.6$ falls below 0 around $\sigma=20$, the MCLE for $p=0.4$ remains positive throughout the range of $\sigma$ shown in the figure. 

\section{The maximum Lyapunov exponent conditioned by infinitely large inputs} \label{section lambda infty}

In this work, our main interest is to determine whether the MCLE falls below 0 with sufficiently amplified inputs. However, it cannot be answered by observing the numerical simulations shown in Figure.\ref{fig:lambda}(a) even if simulations are performed for quite large $\sigma$, which motivates us to derive the analytic value of the MCLE conditioned by infinitely large inputs, $\lambda_\infty$. We introduce a scaling parameter of input signals, $\xi >0$, and then represent the amplified input signals by $\{\xi s(t)\}_t$. The MCLE conditioned by infinitely large inputs is denoted by $\lambda_\infty := \lim_{\xi\to \infty}\lambda$. 

The quantity $\lambda_\infty$ can be derived analytically as follows. We assume that $K(t)$ converges to a certain value, $K_\infty$, as $\xi$ becomes sufficiently large. Whether this assumption is valid or not depends on the nature of the input time series. We will discuss a case where $K(t)$ does not converge in Sec.\ref{discussion}. The value of $K_\infty$ is given by
\begin{eqnarray}
K_\infty =-\alpha g^2+\dfrac{4}{\pi}\alpha g^2 \bigg(\dfrac{\pi}{2}p+(1-p)\arctan{\sqrt{1+\pi K_\infty}}\bigg). \nonumber \\
\label{selfconsistentEq_infty}
\end{eqnarray}
This expression for $K_\infty$ is obtained by replacing $s(t)$ with $\xi s(t)$ in Eq.(\ref{selfconsistentEq}) and considering the limit $\xi \to \infty$. Taking this limit and substituting $K_\infty$ for $K(t)$ in Eq.(\ref{MCLE}), we obtain
\begin{eqnarray}
\lambda_\infty = \dfrac{1}{2} \log \alpha g^2
\bigg( \dfrac{1-p}{\sqrt{1+\pi K_\infty}} \bigg) .
\label{MCLE_infty}
\end{eqnarray}
We have confirmed that the value of $\lambda_\infty$ obtained from Eqs.(\ref{MCLE_infty}) is consistent with the results of numerical simulations (Fig.\ref{fig:lambda}(b)).

Figure \ref{fig:lambda}(b) depicts the relationship between $p$ and $\lambda_\infty$ for various values of the recurrent weight intensity $g$. We define the value of $p$ at $\lambda_\infty=0 $ as the “critical input partiality", $p_c$. Because the condition $\lambda_\infty>0$ always holds for $p<p_c$, we conclude that even sufficiently amplified input signals cannot suppress the chaotic activity of the RNN if $p < p_c$. 

\begin{figure}[t]
    \centering
    \includegraphics[width=0.9\linewidth]{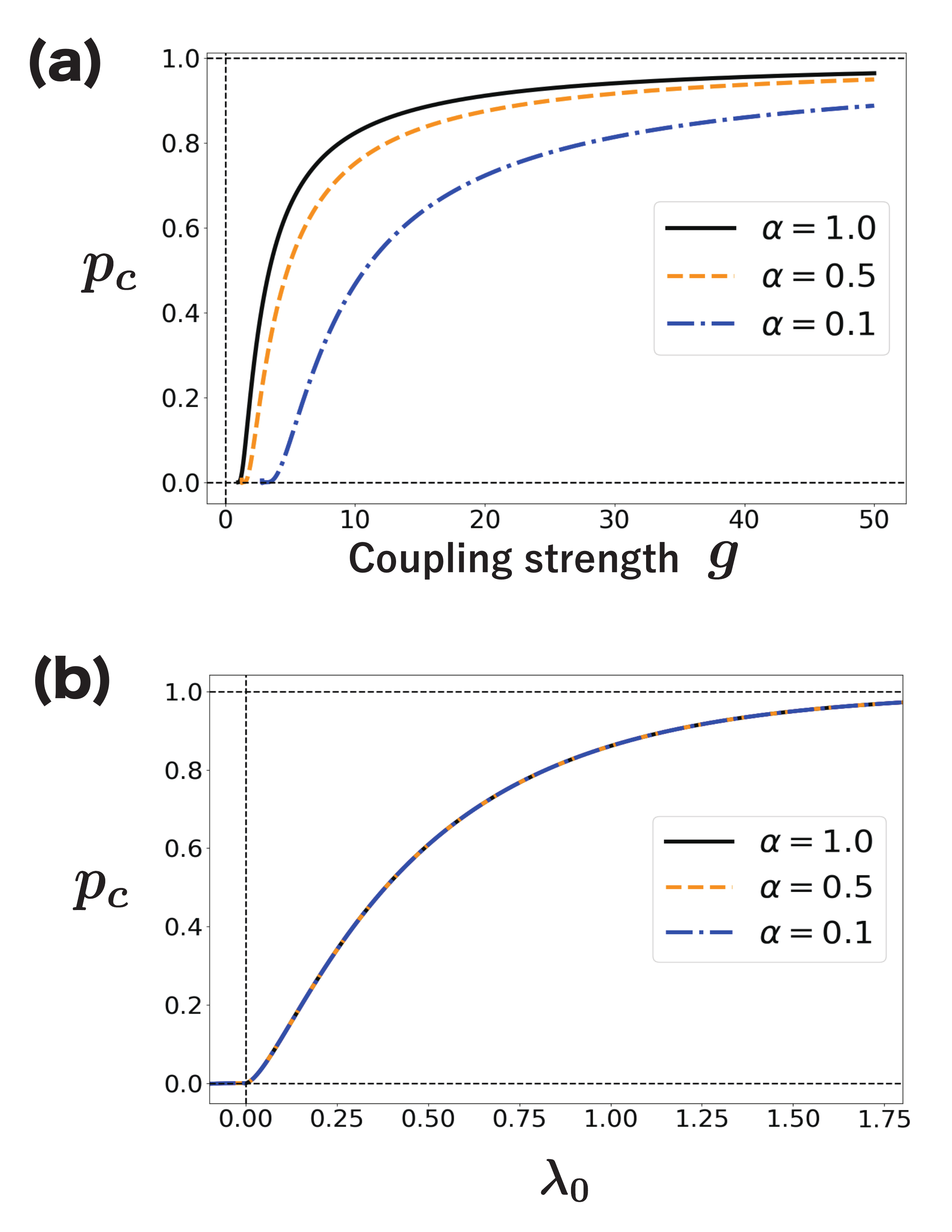}
    \caption{
    (a) Relation between the coupling strength, $g$, and the critical input partiality, $p_c$, given by Eq.(\ref{g_pc_relation}) for several values of the sparsity, $\alpha$.
    (b) The same data as in (a) plotted with respect to $\lambda_0$ rather than $g$. It is seen that when plotted in this manner, the data for $p_c$ fall along the same curve for each value of $\alpha$ considered.
    }
    \label{fig:pc}
\end{figure}

Figure \ref{fig:pc}(a) depicts the dependence of $p_c$ on $g$ with fixed $\alpha$. We obtain the curve by solving Eq.(\ref{MCLE_infty}) for $K_\infty$ with $\lambda_\infty=0$ and substituting this $K_\infty$ into Eq.(\ref{selfconsistentEq_infty}), yielding
\begin{eqnarray}
p_c &&= 1 - \dfrac{1}{\alpha g^2} \bigg[ 1-\pi \alpha g^2 \nonumber \\
    &&\quad + 4 \alpha g^2 \bigg(\dfrac{\pi}{2}p_c + (1-p_c) \arctan{\big((1-p_c) \alpha g^2\big)} \bigg) \bigg]^{1/2}.\nonumber \\ 
\label{g_pc_relation}
\end{eqnarray}
Below the curve, $\lambda_\infty$ is positive, and thus, in this region, chaos is not suppressed no matter how strong the input. As seen in Figure \ref{fig:pc}(a), $p_c$ is an increasing function of $g$. 

We next investigate the effect of sparsity $\alpha$ on $p_c$. Plotting the $g$-$p_c$ curves with several values of $\alpha$, we find that a sparser RNN results in a smaller value of $p_c$ (Fig.\ref{fig:pc}(a)). This is intuitively understandable, because the dynamics of a sparser RNN are less chaotic, and thus a smaller value of the input partiality is sufficient to control the chaos. To take account of this relationship, we introduce the MLE of the RNN with no input, denoted by $\lambda_0$. Clearly, $\lambda_0$ quantifies the strength of chaos in spontaneous activity of the RNN, and it can be determined analytically by substituting $s(t)\equiv0$ into Eq.(\ref{selfconsistentEq}) and Eq.(\ref{MCLE}), yielding 
\begin{eqnarray}
\label{lambda_0_K}
K &&= \alpha g^2 \left( -1 + \dfrac{4}{\pi} \arctan{\sqrt{1+\pi K}} \right),\\
\label{lambda_0}
\lambda_0 &&= \dfrac{1}{2} \log \dfrac{\alpha g^2}{\sqrt{1+\pi K}}.
\end{eqnarray}
Interestingly, we find that when $p_c$ is plotted with respect to $\lambda_0$, the resulting curves for all values of $\alpha$ coincide, as seen in Figure \ref{fig:pc}(b). The reason for this coincidence is easily understood by considering Eqs.(\ref{g_pc_relation})-(\ref{lambda_0}). From Eqs.(\ref{lambda_0_K}) and (\ref{lambda_0}), we see that $\lambda_0$ is a function of $\alpha g^2$. Writing the corresponding inverse function as $\alpha g^2 = f(\lambda_0)$, and substituting this into Eq.(\ref{g_pc_relation}), we obtain $p_c$ expressed as a function of $\lambda_0$ alone. This finding implies that $p_c$ depends primarily on the strength of spontaneous chaos, independently of how sparse the recurrent connection is.

\section{Performance of a partially driven reservoir computing }

Finally, we study the information processing capability of a partial-input RNN employed as a reservoir for RC. Memory capacity \cite{Jaeger2001b} is a commonly used benchmarks for RC. It is a measure of the ability of a reservoir to perform short-term memory tasks through the reconstruction of its past input signals. The memory capacity is defined as follows. From the $N$ neurons, $K$ ($1\leq K \leq N$) lead-out units are randomly chosen, and represented by a vector $\bm{\tilde{x}}(t) \in \mathbb{R}^K$. The reservoir's output is defined as $\hat{z}(t):=\bm{w}^\top \bm{\tilde{x}}(t)$, where the vector $\bm{w}\in \mathbb{R}^K$ represents the output weights. In a $\tau$-delay memory task, the reservoir at time $t$ is required to output the previous input signal $s(t-\tau)$, and the output weights are trained to minimize the mean squared error between $\hat{z}(t)$ and the desired output $s(t-\tau)$. This is accomplished with a least-squares method, and the trained output weights are determined as $\bm{\hat{w}}=(\bm{X} \bm{X}^\top)^{-1}\bm{X} \bm{s}$, where $\bm{X}:=(\bm{\tilde{x}}(1) \cdots \bm{\tilde{x}}(T))$ and $\bm{s}:= (s(1) \cdots s(T))^\top$ ($T$ being the length of the simulation). After training, we evaluate the task performance $M_{\tau}$ defined as
\begin{eqnarray}
M_{\tau} := 1- \dfrac{\langle (\hat{z}(t)-s(t-\tau))^2 \rangle} { \langle s(t)^2 \rangle},
\label{Mtau}
\end{eqnarray}
where the brackets represent the time average. Because the numerator of the second term in Eq.(\ref{Mtau}) is the mean squared error, $M_\tau$ approaches 1 as the reservoir learns to accurately reconstruct its past input $s(t-\tau)$. The memory capacity $MC$ is defined as the sum of the $M_{\tau}$, $MC := \sum_{\tau=1}^{\infty} M_{\tau}$.
It has been mathematically proved that $MC$ satisfies the inequality $0\leq MC \leq K$ \cite{Dambre2012, Jaeger2001b}. 

\begin{figure}[t]
    \centering
    \includegraphics[width=\linewidth]{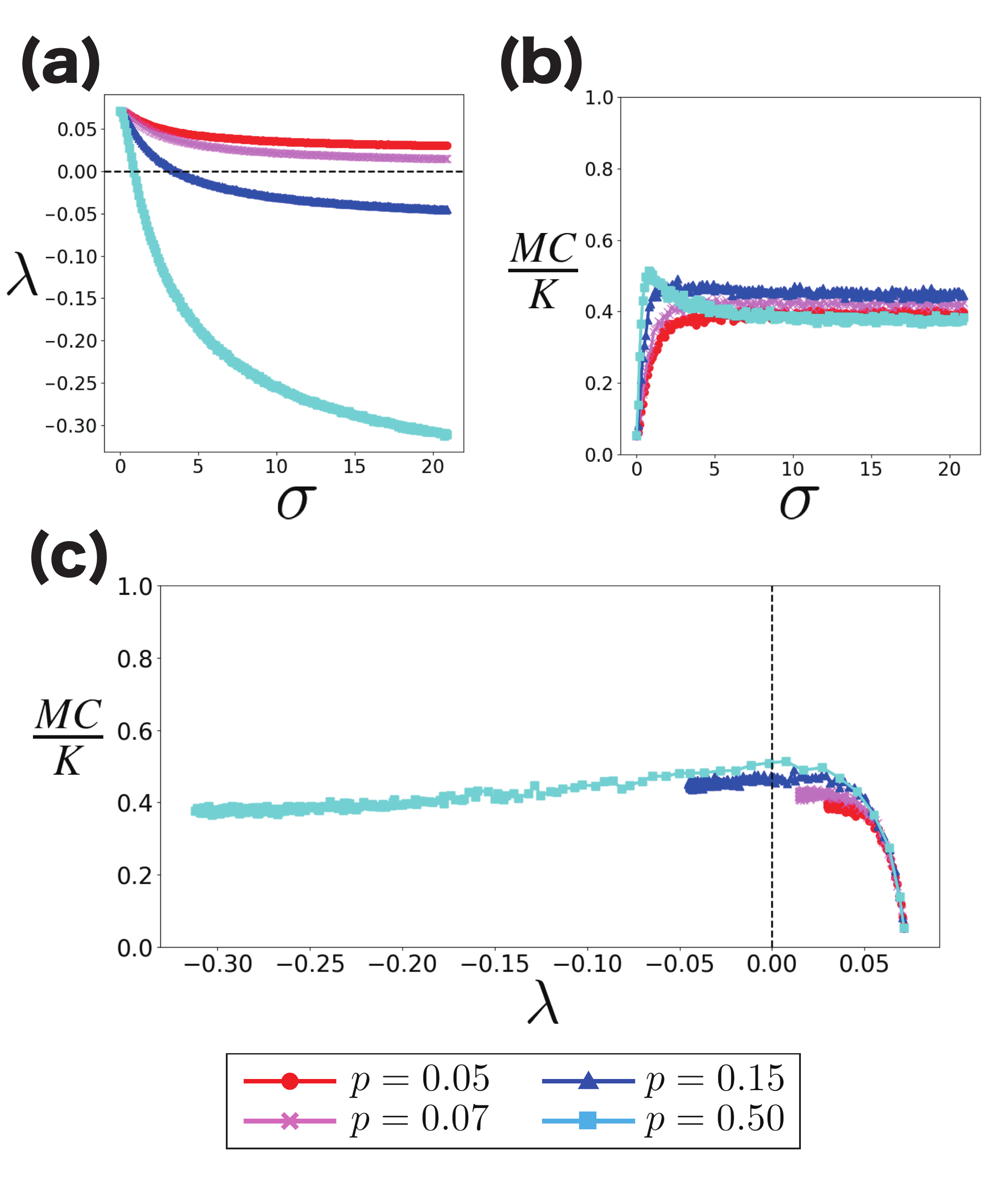}
    \caption{
    Relation between the memory capacity, $MC$, and the maximum conditional Lyapunov exponent (MCLE), $\lambda$, for network size $N=1000$, coupling strength $g=1.5$, sparsity $\alpha=1.0$, and number of lead-out nodes $K=10$. The values of (a) $\lambda$ and (b) $MC$ are respectively plotted as functions of the standard deviation of the input signals, $\sigma$. (c) Each plot represents $(\lambda, MC/K)$ with various values of $\sigma$ ($0.01 \leq \sigma \leq 20$). The sum of the $M_\tau$ is calculated up to $\tau=500$.
    }
    \label{fig:memory_capacity}
\end{figure}

Assuming that the input signal $s(t)$ is Gaussian white noise with zero mean and variance $\sigma^2$, we calculated both $\lambda$ and $MC$. The results are plotted as functions of $\sigma$ in Figure \ref{fig:memory_capacity}(a) and (b). As previously noted, an increase in the input magnitude leads to a decrease in the MCLE (as seen in Fig.\ref{fig:memory_capacity}(a)), with the result that the plot in Fig.\ref{fig:memory_capacity}(c) shifts leftward as $\sigma$ increases. From Eq.(\ref{g_pc_relation}), $p_c$ is found to be approximately $0.074$ under the conditions employed in Fig.\ref{fig:memory_capacity}. When $p=0.15$ and $p=0.50$, the plots intersect the vertical line $\lambda=0$ in Fig.\ref{fig:memory_capacity}(c), as our theory predicts, and the memory capacity reaches its maximum value near $\lambda=0$. Contrastingly, the plots with $p=0.05$ and $p=0.07$ remain in the chaotic domain ($\lambda > 0$), and the memory capacity remains relatively low. It is thus seen that once input connections have been built such that $p$ exceeds $p_c$, optimal computational capability can be realized only by amplifying the input signals appropriately. This finding should be helpful for the physical reservoir computing paradigm, because amplifying input signals is generally easier and more cost effective than adding new input connections.

\section{Discussion} \label{discussion}
In the present work, we have examined a partial-input RNN with rate neurons and have analytically shown the existence of a critical input partiality $p_c$ that determines whether the chaotic activity can be suppressed by input signals. In our theory, RNNs are assumed to have rate-based neurons and recurrent weights sampled from Gaussian distribution. It is our future work whether there exists critical input partiality in other types of an RNN such as that with spiking neurons or heavy tailed recurrent weights \cite{Kusmierz2020}. 

In the derivation of $\lambda_\infty$ (Sec.\ref{section lambda infty}), we have assumed $\lim_{\xi \to \infty} K(t)$ to exist. However, there are some examples where this assumption does not hold. For example, let us consider the case where an input time series, $s(t)$, has a non-negligible number of zeros. Then, even if the scaling parameter $\xi$ approaches infinity, $\xi s(t)$ also has a non-negligible number of zeros, and thus, the value of $K(t)$ determined by Eq.(\ref{selfconsistentEq}) does not converge. Although it remains to be seen what condition on input signals is required for the existence of $\lim_{\xi \to \infty} K(t)$, we believe that our theory holds for a wider variety of input signals than Gaussian white noise. 

We have focused on the sign of an MCLE to investigate the echo state property (ESP) of an RNN. In RNNs, the sign of the MCLE has been widely regarded as a representative indicator of whether the ESP holds or not. However, it should be noted that a negative MCLE does not necessarily guarantee that the ESP holds. We discuss two typical cases below.

Firstly, a negative MCLE does not always guarantee ESP. This distinction stems from the fact that the Lyapunov exponent primarily characterizes local stability, whereas ESP is related to the global stability of the network’s response to identical input signals. For example, if there are two locally stable attractors, the different initial states in the reservoir can lead to convergence to different attractors. This scenario results in different outputs in response to identical inputs, violating the conditions of ESP.

Moreover, there is another counterexample where an RNN with a positive $\lambda_\infty$ can behave as having a stable attractor under certain tuned input signals. This can be realized by inputs that are precisely constructed by chaos control methods, such as those used in the Poincaré map \cite{Ott1990}. However, in the context of reservoir computing, chaos control conditions are rarely satisfied in realistic situations because the input signals are typically predetermined as training data.

Taking all of the above into consideration, a negative MCLE is a reliable, though not infallible, indicator of ESP. In fact, our numerical simulations demonstrated that the ESP always holds with the negative MCLE (data not shown). Consequently, we believe that these exceptions does not substantially affect the general applicability of our results.

We have confirmed that memory capacity is maximized near the critical value of MCLE, $\lambda=0$, which corresponds to the “edge of chaos." Our theory suggests that we can readily construct a partial-input RNN at the edge of chaos by tuning the magnitude of input signals, as long as the input partiality exceeds $p_c$. The present study provides a possible novel approach to designing reservoir computing.

\begin{acknowledgments}
S.T. was supported by JSPS KAKENHI Grant No. JP22J21559. T.A. was supported by JSPS KAKENHI Grant No.JP20K21810, No.JP20H04144, and No.JP20K20520.
\end{acknowledgments}


\providecommand{\noopsort}[1]{}\providecommand{\singleletter}[1]{#1}%

\end{document}